# Spectroscopic identification of 2D conjugated polyphthalocyanines

Vitaly I. Korepanov, Daria M. Sedlovets *

Institute of microelectronics technology and high purity materials RAS

*e-mail*: sedlovets@iptm.ru

**Abstract**

Since the discovery of polyphthalocyanines (PPCs) in late 1950s, numerous attempts have been made to synthesize this 2D polymer by different approaches. Interestingly, the reported IR, Raman and UV-vis spectra of PPCs show drastic variation depending on the synthesis conditions. In this work, we show that the spectral data obtained in some works should be assigned not to the target polymer, but rather to octacyano phthalocyanine (OCP), which is an early step of the reaction. We discuss the spectral signatures of the well-polymerized and monomeric PCs based on reliable experimental data and support the spectral assignments with DFT calculations.

**Introduction**

Attempts to synthesize polymeric phthalocyanines by the reaction of pyromellitic tetranitrile (PMTN) with metals date back to late 1950s[1,2]. Since then, the reaction has been studied in many works (see, for example, reviews [3,4] and references therein). It has been shown, in particular, that depending on temperature, the reaction can proceed in two ways: at about 200°C the major product is octacyano phthalocyanine (OCP), while the 2D polymerization requires significantly higher temperatures of above 350°C[3,5–9] (fig. 1).

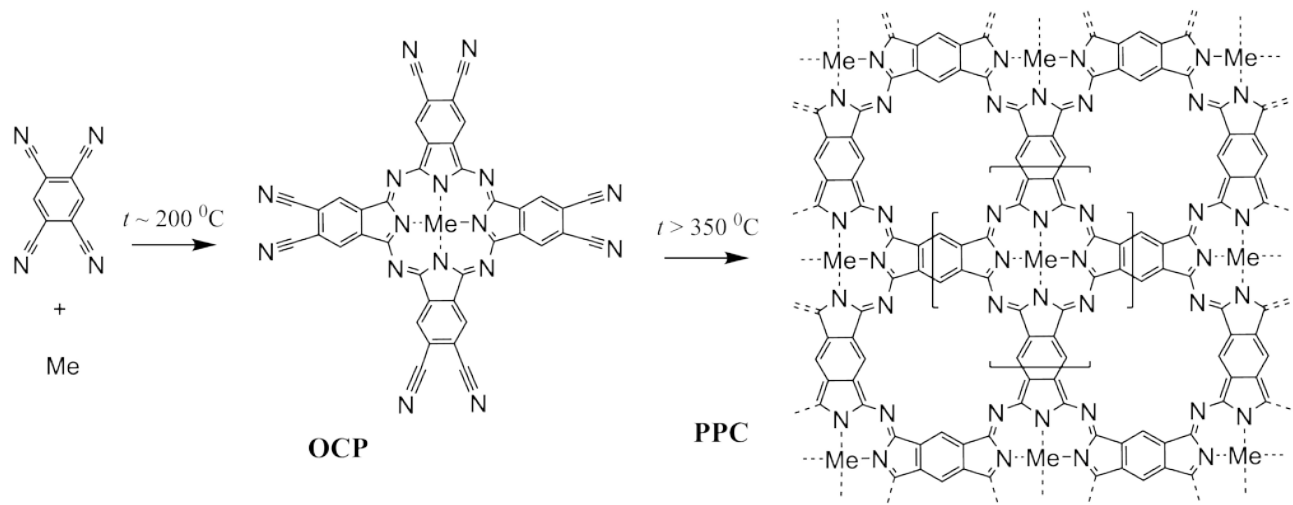

Figure 1. Reaction products of pyromellitic tetranitrile (PMTM) with metals according to [3,5–9].

In a recent wok, however, the authors claimed the synthesis of conjugated cobalt polyphthalocyanine (CoPPC) by microwave heating in pentanol at 180°C[10]. The material was characterized by IR, Raman and UV-vis spectroscopy, and studied as a catalyst for flexible Li–$CO_2$ batteries. Herein, we do not discuss unique properties of the synthesized material as a catalyst, but it should be noted that in both polymeric and monomeric PCs the metal atom has similar surrounding, and is therefore likely to have similar catalytic properties. In the present work, we address the spectral identification of the reaction products. In particular, we show that the synthesized product cannot be interpreted as a polymer, while all spectra are in agreement with OCP as the reaction product.

To address the correct spectral patterns of well-polymerized conjugated PCs, we consider CuPPC as a model compound. It has been shown that different transition metal PCs within the row Fe-Co-Ni-Cu-Zn have very close spectroscopic properties[11]. Moreover, in the recent experimental study, the metal-free PPC ($H_2$PPC) was also found to have a similar spectral patterns in IR, Raman and UV-vis[12]. Vibrational bands observed in the discussed spectral regions belong to the intramolecular modes of the organic skeleton; the electronic transitions in the visible region also belong mostly to the conjugated structure of π-electrons. The general spectral assignment therefore is applicable to the whole series.

**IR spectra**

Both monomeric and polymeric PCs have highly symmetric molecular structures ($D_{4h}$ and 4/mmm point group correspondingly) and are composed of similar fragments. The key difference however is that while the PC has 57 atoms in the molecule, the translational unit in the polymer has only 33 atoms (fig. 1). Direct consequence of this fact is that the vibrational spectra of PPCs should show much less bands than those of the monomeric form. Taking into account the inversion symmetry, each vibrational band should be active only in Raman or IR, but not in both. As a result, the predicted spectral pattern of CuPPC is rather simple (fig. 2). Experimental spectrum of well-polymerized CuPPC also has a small number of bands; the overall agreement between calculated and experimental spectra in terms of band positions and relative intensities is very good.

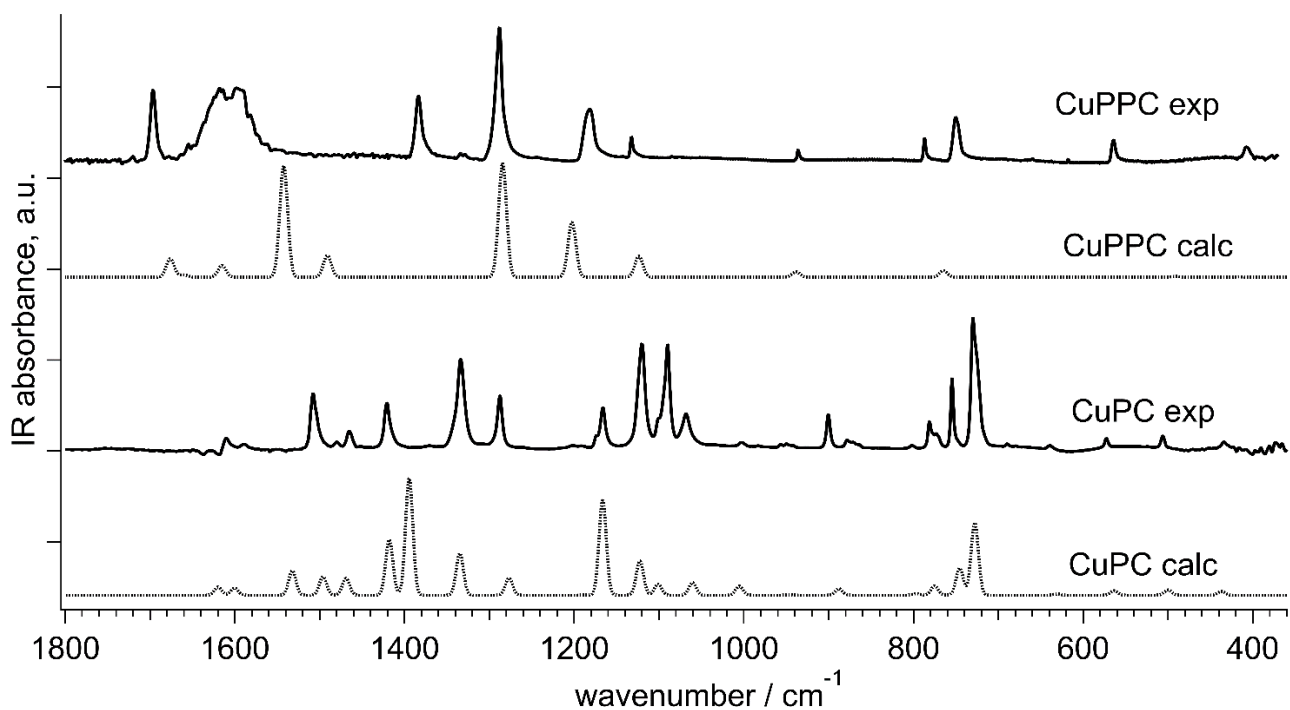

Figure 2. Comparison of infrared spectra of polymeric and monomeric phthalocyanines. Solid lines: experimental spectra (top: CuPPC, bottom: CuPC), underlying dotted lines are the corresponding calculated spectra.

Important quantitative signature, which can help to reliably distinguish the OCP from the polymer is the relative intensity of the C≡N stretching band at ~2220 cm$^{-1}$. The alleged CoPPC reported in the work [10] shows this band with a significant intensity (fig. 1d in [10]); the overall spectral pattern is in a close agreement with the OCP spectra reported in [13]. For the well-polymerized CuPPC, the band in this spectral region was not observed[14] (not shown in fig. 2).

The spectral identification of the material obtained in the work[10] at 180°C is therefore quite straightforward, except for the bands at around 1750 cm$^{-1}$. The latter come from the carbonyl groups, and are probably the result of oxidation of the material with residual oxygen. In our experiments, we also observed the carbonyl bands in IR spectra unless we completely removed oxygen from the reaction media (e.g. by purging inert gas or hydrogen). These bands are an indication of an undesired by-product.

**Raman spectra**

In Raman spectra, the main difference between the monomer and the polymer is that the former latter shows several close-lying intense bands in 1300-1700 cm$^{-1}$ range, while the former shows a characteristic spectrum with many bands of comparable intensity. Both DFT calculations and experiment agree in this description (fig. 3). The bands in the spectrum of the polymer belong to the system of conjugated CC and CN bonds; similar spectral pattern was observed for $H_2$PPC[12].

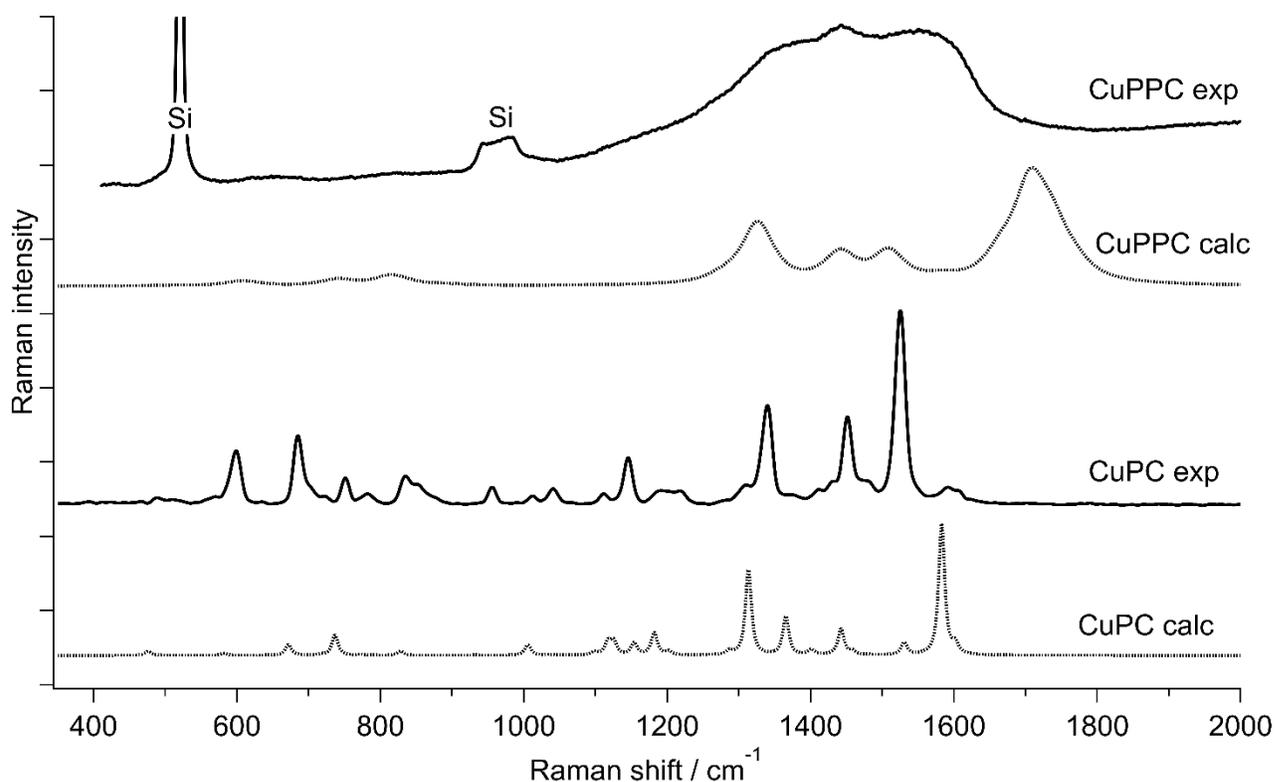

Figure 3. Comparison of Raman spectra of polymeric and monomeric phthalocyanines. Solid lines: experimental spectra (top: CuPPC, bottom: CuPC), underlying dotted lines are the corresponding calculated spectra. The silicon waver bands are marked in the top curve.

**Optical absorption**

In monomeric PCs, the electronic conjugation spreads mainly over the 18 π electrons of the inner ring[15], and the characteristic signature of such structure is the Q band in 650-700 nm range, attributed to the HOMO-LUMO transition. Electronic structure of the polymeric PCs however is drastically different: depending on the central metal they either are semi-metals or have a low band gap in IR spectral region [14]. Therefore, the Q band in visible region should not be observed in polymers. The spectrum of well-polymerized PC is analogous to that of another 2D conjugated π-system, graphene, i.e. shows continuous absorption throughout the optical region (fig. 5).

Quantum-chemical calculations confirm this conclusion. Both HOCO and LUCO of MePPCs belong to the conjugated π-electronic system (fig. 4); the energy difference between them is 0.02 eV as compared to ~1.9 eV for CuPPC.

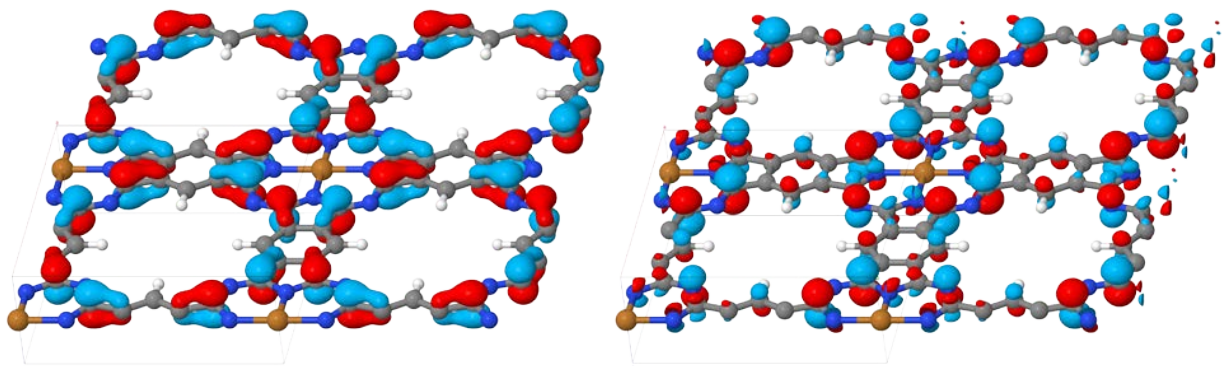

Figure 4. Crystal orbitals of MePPCs: HOCO (left) and LUCO (right). The calculated energy difference between these two orbitals is 0.02 eV.

The UV-vis absorption spectra of PPCs have been discussed in several works [3,6,7,13,16,17]. Typically, the material was dissolved for the measurements. It was shown that concentrated sulfuric acid can dissolve some portion of the polymer due to protonation[13]. Nevertheless, the material of high polymerization degree was found to be insoluble even in this solvent[14] showing thus the behavior similar to graphene. The alleged solubility of CoPPC in dimethyl sulfoxide (DMSO) and N-methylpyrrolidone (NMP) reported in the work [10] is a clear indication of low molecular weight. In agreement with the FTIR data the material should be assigned to OCP.

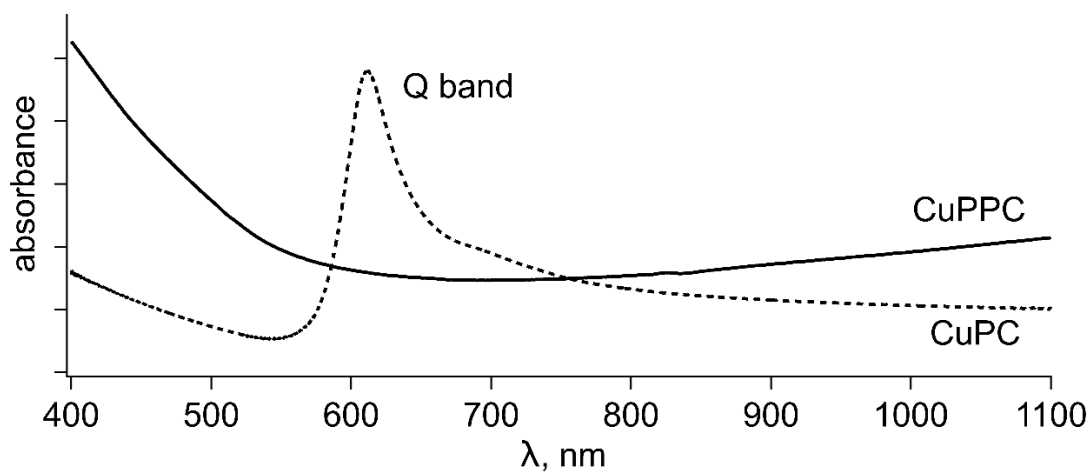

Figure 5. Comparison of UV-vis absorption spectra of polymeric and monomeric phthalocyanines in thin films on fused silica. Solid line: CuPPC, dotted line: CuPC.

The correct comparison of the absorption spectra of monomeric and polymeric phthalocyanines should be done within the same aggregation state. Fig. 5 shows optical of thin films of CuPPC and CuPC in the visible range; as expected, the Q band is observed only in the spectrum of the monomer.

**Computation details**

DFT calculations of geometry and vibrational spectra of the polymer were done with Quantum Espresso package[18]. The PBEsol functional was taken with high-throughput ultrasoft pseudopotentials[19] with cut-off for wavefunctions/charge density of 50/320 Ry. 4x4x4 grid of k-points was used. For the copper atom, in order to take an advantage of the closed-spin electronic system, the pseudopotential of zinc was assumed; as was discussed above, the applicability of this approximation has been well tested in the literature [11]. Raman tensor was calculated with norm-conserving pseudopotentials[20] with the cut-off of 88/352 Ry. Since the PBEsol functional typically underestimates vibrational frequencies by few percent[21], for comparison with the experimental data, the computed vibrational frequencies were corrected with a single scaling factor of 1.09.
Crystal orbitals were visualized with Jmol[22]

Calculations of the monomeric phthalocyanine are the courtesy of Dr. Yury V. Vishnevskiy (Universität Bielefeld). The Turbomole package was used. The BP86 GGA functional was taken with def2-SV(P) basis set.

**Experiment**

The synthesis of CuPPC was done according to the previously reported procedure[12,14]. Briefly, a thin layer (few nm) of copper metal was RF-sputtered on a substrate and placed in a CVD reactor. Therein, it was exposed to the PMTN vapor at 420°C for a few hours. As a substrate, KBr plate was used (for IR measurements) or fused silica (for UV-vis spectra) or silicon wafer (for Raman spectroscopy). FTIR spectra were taken with a Bruker IFS-113v spectrometer in the 400-4000 cm$^{-1}$ range under 1 cm$^{-1}$ resolution. Optical measurements were made with a Specord-50 spectrometer and Bruker Vertex 70 V spectrometer. Raman spectra were measured with a Bruker Senterra micro-Raman system under 532 nm excitation and laser power 2 mW.

**Conclusions**

In this work, we report spectroscopic signatures of 2D conjugated polymeric phthalocyanines. We show that IR, Raman and UV-vis spectra of the polymer are significantly different from those of the monomeric phthalocyanines. The spectral assignment is supported by DFT calculations. We hope that the provided data will assist the reliable identification of 2D PPCs.

**Acknowledgements**

We acknowledge Dr. Yury V. Vishnevskiy (Universität Bielefeld) for calculations of monomeric phthalocyanine. This work was supported by Russian Science Foundation (project no. 17-73-10128).

**Supporting Information**

Structure of the unit cell of CuPPC used in DFT calculations with Quantum Espresso.

```
ibrav=6,
celldm(1)=20.2153992727694, ! (Bohr)
celldm(3)=0.35, ! (in units of celldm(1))

ATOMIC_POSITIONS alat ! (in units of celldm(1))
N      0.000000000  0.185548616  0.000000000
N      0.185548616  0.000000000  0.000000000
N      0.223215250  0.223215250  0.000000000
N      0.776779151  0.223215250  0.000000000
N      0.814445782  0.000000000  0.000000000
N      0.223215250  0.776779151  0.000000000
N      0.000000000  0.814445782  0.000000000
N      0.776779151  0.776779151  0.000000000
H      0.499997201  0.761843302  0.000000000
H      0.761843302  0.499997201  0.000000000
H      0.499997201  0.238151099  0.000000000
H      0.238151099  0.499997201  0.000000000
C      0.104749470  0.259787305  0.000000000
C      0.259787305  0.104749470  0.000000000
C      0.895244932  0.259787305  0.000000000
C      0.609918304  0.066362504  0.000000000
C      0.740207095  0.104749470  0.000000000
C      0.499997201  0.135635809  0.000000000
C      0.390076098  0.066362504  0.000000000
C      0.259787305  0.895244932  0.000000000
C      0.066362504  0.609918304  0.000000000
C      0.104749470  0.740207095  0.000000000
C      0.135635809  0.499997201  0.000000000
C      0.066362504  0.390076098  0.000000000
C      0.609918304  0.933631897  0.000000000
C      0.933631897  0.609918304  0.000000000
C      0.740207095  0.895244932  0.000000000
C      0.895244932  0.740207095  0.000000000
C      0.499997201  0.864358590  0.000000000
C      0.864358590  0.499997201  0.000000000
C      0.390076098  0.933631897  0.000000000
C      0.933631897  0.390076098  0.000000000
Cu     0.000000000  0.000000000  0.000000000
```